\documentclass[a4paper, 12pt]{iopart}

\usepackage{graphicx}
\usepackage{bm}
\usepackage{epsfig}
\usepackage{amssymb}
\usepackage{amsfonts}
\usepackage{hyperref}

\newcommand{\eV}{{\rm eV}}

\newcommand{\TeV}{{\rm TeV}}

\newcommand{\Mpc}{{\rm Mpc}}

\newcommand{\cm}{{\rm cm}}

\newcommand{\s}{{\rm s}}

\newcommand{\sr}{{\rm sr}}

\newcommand{\Mpl}{M_{\rm Pl}}
\newcommand{\nubar}{\bar{\nu}}

\graphicspath{{PLOTS/}}

\begin{document}

\hspace{11cm}{DESY 09-168}

\title{Possible cosmogenic neutrino constraints on Planck-scale Lorentz violation}

\author{David M.~Mattingly$^{1}$, Luca Maccione$^{2}$, Matteo Galaverni$^{3}$, Stefano Liberati$^{4,5}$, G\"unter Sigl$^{6}$}

\address{
$^{1}$ University of New Hampshire, Durham, NH 03824\\
$^{2}$ DESY, Theory Group, Notkestra\ss e 85, D-22607 Hamburg, Germany \\
$^{3}$ INAF-IASF Bologna, Via Gobetti 101, I-40129 Bologna, Italy \\
$^{4}$ SISSA, Via Beirut, 2-4, I-34151, Trieste, Italy \\
$^{5}$ INFN, Sezione di Trieste, Via Valerio, 2, I-34127, Trieste, Italy \\
$^{6}$ {II}. Institut f\"ur Theoretische Physik, Universit\"at Hamburg, Luruper Chaussee 149, D-22761 Hamburg, Germany
}

\eads{
\mailto{davidmmattingly@comcast.net},
\mailto{luca.maccione@desy.de},
\mailto{galaverni@iasf.bo.it},
\mailto{liberati@sissa.it},
\mailto{guenter.sigl@desy.de}
}

\begin{abstract}
We study, within an effective field theory framework, $O(E^{2}/\Mpl^{2})$ Planck-scale suppressed Lorentz invariance violation (LV) effects in the neutrino sector, whose size we parameterize by a dimensionless parameter $\eta_{\nu}$. We find deviations from predictions of Lorentz invariant physics in the cosmogenic neutrino spectrum. For positive $O(1)$ coefficients no neutrino will survive above $10^{19}~\eV$. The existence of this cutoff generates a bump in the neutrino spectrum at energies of $10^{17}~\eV$. Although at present no constraint can be cast, as current experiments do not have enough sensitivity to detect ultra-high-energy neutrinos, we show that experiments in construction or being planned have the potential to cast limits as strong as $\eta_{\nu} \lesssim 10^{-4}$ on the neutrino LV parameter, depending on how LV is distributed among neutrino mass states. Constraints on $\eta_{\nu} < 0$ can in principle be obtained with this strategy, but they require a more detailed modeling of how LV affects the neutrino sector.
\end{abstract}

\section{Introduction}

Over the last fifteen years there has been consistent
theoretical interest in possible small deviations from the exact
local Lorentz Invariance (LI) of general relativity as well as a
flourishing of observational tests. The theoretical interest is
driven primarily by ideas in the Quantum Gravity (QG) community that
Lorentz invariance may not be an exact local symmetry of the vacuum.
The possibility of outright Lorentz symmetry violation (LV) or a
different realization of the symmetry than in special relativity has
arisen in string theory~\cite{KS89, Ellis:2008gg}, Loop
QG~\cite{LoopQG,Rovelli:2002vp,Alfaro:2002ya}, non-commutative
geometry~\cite{Carroll:2001ws, Lukierski:1993wx,
AmelinoCamelia:1999pm, Chaichian:2004za}, space-time
foam~\cite{AmelinoCamelia:1997gz}, some brane-world
backgrounds~\cite{Burgess:2002tb}, and condensed matter analogues of
``emergent gravity''~\cite{Analogues}. 
As well, allowing the fundamental theory of
gravity to be non-relativistic in the ultraviolet can make gravity
renormalizable while avoiding some of the other pathologies that
plague renormalizable gravitational actions with higher derivative
terms~\cite{Horava:2009uw}.

Constructing useful tests for the various active models and ideas is
therefore vital.  Since there are so many theoretical models around,
a good approach is to work within a calculable framework where all
possible LV terms are parameterized.  Each theory then picks out a
certain combination of terms which can be constrained. 
In this vein, a standard method is to simply analyze a
Lagrangian containing the standard model fields and all LV operators
of interest that can be constructed by coupling the standard model
fields to new LV tensor fields that have non-zero vacuum expectation
values\footnote{There are other approaches to either violate or
modify Lorentz invariance, that do not necessarily yield a low
energy EFT (see \cite{AmelinoCamelia:2008qg} and refs therein).
However, these models do not easily lend themselves to particle
physics constraints as the dynamics of particles is less well
understood and hence we do not consider them here. In particular, we remark here that ideas of deformation, rather than breaking, of the Lorentz symmetry (see, e.g., \cite{AmelinoCamelia:2000mn}) do not have an ordinary-EFT formulation, hence they cannot be tested with the arguments presented in this work.}. 

All renormalizable LV operators that can be added to the standard model
in this way are known as the Standard Model Extension
(SME)~\cite{Colladay:1998fq}. These operators all have mass
dimension three or four and can be further classified by their
behavior under CPT. Higher mass dimension operators can be systematically explored as well, which is useful in case the naive EFT hierarchy breaks down due to other new physics (e.g.~SUSY) or quantum gravity introducing a custodial mechanism for the renormalizable operators in the infrared. 

The CPT odd dimension five kinetic terms for QED
coupled to a non-zero vector field were written down
in~\cite{Myers:2003fd} while the full set of dimension five
operators with a vector were analyzed in~\cite{Bolokhov:2007yc}. The
dimension five and six CPT even kinetic terms for QED for particles
coupled to a non-zero background vector, which we are primarily
interested in here, were partially analyzed
in~\cite{Mattingly:2008pw}.  The full set of dimension five and six
operators for QED has recently been introduced in~\cite{Kostelecky:2009zp}.
It is notable that SUSY forbids renormalizable operators for matter
coupled to non-zero vectors~\cite{GrootNibbelink:2004za} but permits
certain nonrenormalizable operators at mass dimension five and six.

Many of the parameterized LV operators have been very tightly
constrained via direct observations (see~\cite{Mattingly:2005re,Jacobson:2005bg, AmelinoCamelia:2008qg,
Liberati:2009pf} for reviews). In particular, the dimension five and six CPT even operators
for LV with a vector field have been recently
directly\footnote{Notice that all these operators can be indirectly
constrained by EFT arguments~\cite{Collins:2004bp}, as higher
dimension LV operators induce large renormalizable ones if we assume
no other relevant physics enters between the TeV and $\Mpl$
energies.  SUSY, however, is an example of new relevant physics that
can change this.} constrained in the hadronic sector by exploiting
ultra-high energy cosmic-ray observations \cite{Maccione:2009ju} performed by the 
Pierre Auger Observatory (PAO).  
Indeed, the construction and successful operation of this instrument
has brought UHECRs to the interest of a wide community of scientists and it is expected to allow, in the near future, the assessment of several problems of UHECR physics and also to test fundamental physics (in particular Lorentz invariance in the QED sector) with unprecedented precision
\cite{Galaverni:2007tq,Maccione:2008iw,Galaverni:2008yj}.

The UHECR constraints \cite{Kifune:1999ex,Aloisio:2000cm,AmelinoCamelia:2000zs,Stecker:2004xm,GonzalezMestres:2009di,Scully:2008jp,Maccione:2009ju,Stecker:2009hj} rely on the behavior of particle reaction thresholds with LV, which are one of the best methods in the EFT
approach to constrain nonrenormalizable LV operators.  Many LV
operators give modified dispersion relations for free particles,
where the energy as a function of momentum deviates slightly from the
special relativistic form. For threshold reactions what matters is
not the size of the LV correction to the energy compared to the
absolute energy of the particle, but instead the size of the LV
correction to the mass of the particles in the reaction.  Hence the
LV terms usually become important when their size becomes comparable to the mass of
the heaviest particle.  If the LV term scales with energy as $E^n$,
then this critical energy is $E_{cr} \sim
\left(m^{2}\Mpl^{n-2}\right)^{1/n}$ \cite{Jacobson:2002hd}.
According to this reasoning, the larger the particle mass the higher
is the energy at which threshold LV effects come into play. This is
why $\gtrsim \TeV$ electrons and positrons, but not protons, can be
used to constrain $n=3$ LV \cite{Maccione:2007yc}, and why UHE
protons are needed to obtain constraints on hadronic LV with $n=4$
scaling (which corresponds to CPT even mass dimension five and six
operators).

From this point of view, neutrinos, with their tiny mass of order
$m_{\nu} \simeq 0.01~\eV$ \cite{pdg}, are in principle the most suited
particles to provide strong constraints on LV, at least for
reactions involving {\em only} neutrinos.  One such reaction is that of
neutrino oscillation.  Indeed, for a decade neutrino oscillations
have proven to be excellent tests of the SME and other LV
models~\cite{Coleman:1998ti,Kostelecky:2009zp,GonzalezGarcia:2005xw,Diaz:2009qk,Yang:2009ge}
as when the LV corrections are near the neutrino mass, the
oscillation pattern can change as a function of energy, direction and mass. ICECUBE may even be able to probe dimension six operators with time of flight techniques with TeV neutrinos from distant Gamma Ray Bursts~\cite{GonzalezGarcia:2006na}. UHECR experiments have also the capability of placing constraints on SME parameters by exploiting neutrino oscillations \cite{Bhattacharya:2009tx}\footnote{We notice that in the same work \cite{Bhattacharya:2009tx} the process of neutrino decay is discussed, but in a different context than what we consider here.}.
One can construct complementary and in some measures even more
sensitive neutrino tests of higher dimension operators  
by leveraging observables
which deviate more strongly from their special relativistic values
as the neutrino travel distance increases. One such observable is found to be related to UHE neutrino spectrum observations.

Despite the threshold being low for LV effects to kick in, neutrinos with ultra-high energy are necessary to achieve a signal, as they interact so weakly that the phase space for a LV reaction must be huge to generate an appreciable rate. This requirement implies that, since the LV terms and hence the phase space grow with energy, very large energies are needed. Indeed, for renormalizable operators, where the phase space does not grow quickly enough, reactions of the type we consider here never achieve the necessary rate given current bounds on their coefficients, even for UHE neutrinos over cosmological distances~\cite{Coleman:1997xq,Coleman:1998ti}. However, for the CPT even non-renormalizable operators the situation is different, and we find that for energies of order $10^{17}~\eV$ there can be a significant modification of the spectrum. Next generation neutrino detectors such as ANITA \cite{Gorham:2008yk} and SuperEUSO \cite{Petrolini:2009cg,Santangelo:2009et} are sensitive to neutrinos of energies $>10^{19}~\eV$. Further experiments, like the planned ARIANNA \cite{Barwick:2006tg,Barwick:2009zz} and IceRay \cite{Allison:2009rz}, will cover the range $10^{17}\div 10^{20}~\eV$. 

In the present work we study how limits on the absolute scale of
non-renormalizable CPT even LV neutrino parameters can be obtained
from UHE neutrino observations.  In particular we will determine the energy scale where the neutrino spectrum might begin to deviate as a function of the size of the LV coefficients.  With three neutrino species as well as assorted light leptons possibly involved, the spectrum as a function of various LV parameters can only be computed by a detailed parameterized numerical search.  At this stage, where LV in this sector is only speculative, we feel that such a search is unwarranted.  However, we will show the ``best case'' scenario for a LV neutrino signal with
dimension six operators as well as a scenario where a signal in the
UHE spectrum is much more subtle even though LV is still relatively
strong.

This paper is structured as follows. In section~\ref{sec:theo} we describe the theoretical LV framework in which we will derive the LV effects on the UHE neutrino spectrum. In sections~\ref{sec:nuLV} and \ref{sec:nusplitting} we will give general information on the standard understanding of UHE neutrino generation by UHECRs, and we will present and detail the main LV reactions possibly affecting their spectrum. Section~\ref{sec:bestcase} is devoted to present results in our test cases. In Section \ref{sec:otherprocesses} we discuss the possible role of other processes than neutrino splitting.  In section~\ref{sec:conclusions} we will report our final remarks and conclusions.

\section{Theoretical framework}
\label{sec:theo}

In order to study the phenomenological consequences of LV induced by
QG, the existence of a dynamical framework in which to compute
reaction thresholds and rates is essential.  We assume that the low
energy effects of LV can be parameterized in terms of a local EFT.
Furthermore, for simplicity we assume that only boost invariance is
broken, while rotations are preserved (see \cite{Mattingly:2005re}
for further comments on rotation breaking in this context).
Therefore we introduce LV by coupling standard model fields to a
non-zero vector. More complicated forms could of course be chosen,
however this would introduce significant direction dependence. With
low statistics already for neutrino observatories, disentangling
direction dependence would be a difficult task. 

We focus on the CPT even mass dimension five and six operators
involving a vector field $u^{a}$ (which we assume to describe
the preferred reference frame in which the CMB is seen as isotropic)
coupled to a Dirac neutrino $\psi$ in a mass eigenstate with mass
$m$ that are quadratic in matter fields and hence modify the free
field equations. Neglecting the left-handed helicity of neutrinos
for a moment, the Lagrangian for a generic fermion is the usual
Dirac term plus \cite{Mattingly:2008pw}
\begin{eqnarray} \label{eq:actionfermion}
\overline {\psi}\bigg{[} - \frac {1} {\Mpl} (u \cdot D)^2
(\alpha^{(5)}_L P_L  + \alpha^{(5)}_R P_R) \\
\nonumber - \frac {i} {\Mpl^2} (u \cdot D)^3 (u \cdot \gamma)
(\alpha^{(6)}_L P_L +
\alpha^{(6)}_R P_R)  \\
\nonumber - \frac {i} {\Mpl^2} (u \cdot D) \Box (u \cdot \gamma)
(\tilde{\alpha}^{(6)}_L P_L + \tilde{\alpha}^{(6)}_R P_R) \bigg{]}
\psi~,
\end{eqnarray}
where $u^a$ is a timelike unit vector describing the preferred
frame, $P_R$ and $P_L$ are the usual right and left projection
operators, $P_{R,L}=(1 \pm \gamma^5)/2$, and $D$ is the gauge
covariant derivative. The $\alpha$ coefficients are dimensionless.

For fermions, at $E\gg m$ the helicity eigenstates are almost
chiral, with mixing due to the particle mass and the dimension five
operators. Since we will be interested in high energy states, we
re-label the $\alpha$ coefficients by helicity, i.e.~$\alpha^{(d)}_{+}=\alpha^{(d)}_{R},
\alpha^{(d)}_{-}=\alpha^{(d)}_{L}$. The resulting high energy
dispersion relation for positive and negative helicity particles can
easily be seen from (\ref{eq:actionfermion}) to involve only the
appropriate $\alpha^{(d)}_{+}$ or $\alpha^{(d)}_{-}$ terms. For
compactness, we denote the helicity based dispersion by
$\alpha^{(d)}_{\pm}$. Therefore at high energies we have the
dispersion relation
\begin{equation} 
\label{eq:dispfermionhighE}
E^2 =p^2+m^2 + f^{(4)}_{\pm} p^2 +f^{(6)}_{\pm} \frac{p^4} {\Mpl^2}~,
\end{equation}
where $f^{(4)}_{\pm}=\frac {m} {\Mpl} (\alpha^{(5)}_-  +
\alpha^{(5)}_+) $ and $f^{(6)}_{\pm}= 2\alpha^{(6)}_{\pm} +
\alpha^{(5)}_- \alpha^{(5)}_+$.  We have dropped the
$\tilde{\alpha}^{(6)}_{R,L}$ terms as the $\Box$ operator present in
these terms makes the correction to the equations of motion
proportional to $m^2$ and hence tiny.

The dimension five fermion operators induce two corrections, one
proportional to $E^4$ and one corresponding to a change in the
limiting speed of the fermion away from $c$. The second effect,
generated by $f^{(4)}$ is naturally of order $10^{-30}$ due to the
mass suppression for neutrinos and hence can also be disregarded.
Furthermore, we can drop the positive helicity coefficients, as
the dominant signal will be from left-handed neutrinos produced by
standard model couplings. Finally, since CPT is conserved, and
neutrinos and antineutrinos exist only in opposite states of
helicity, $f^{(6)}_\nu \equiv \eta_\nu = f^{(6)}_{\nubar}$. All
these lovely properties of neutrinos mean we have just one LV
coefficient, while in other cases \cite{Maccione:2009ju} further assumptions, such
as parity preservation, are needed to reduce the number of free
parameters. The final dispersion relation we assume in this work for
each mass eigenstate neutrinos and antineutrinos is
\begin{equation}
\label{eq:finaldisp} 
E_\nu^2=p^2  + m_\nu^2 +  \eta_{\nu I} \frac{p^4}{\Mpl^2}~,
\end{equation}
where $I$ denotes the mass eigenstate of the neutrino.  There is no
reason why each mass eigenstate needs have the same coefficient, and
indeed whether or not they do makes a dramatic difference in the
observed spectrum. In particular, we will study two cases: in the ``flavor blind'' case all the $\eta_{\nu I}$ have roughly the same magnitude, which translates in all the LV effects being essentially independent of flavor; in the ``flavor dependent'' case the $\eta_{\nu I}$ are instead different for different mass states, which makes the LV effects depend upon flavor states. Among the ``flavor blind'' cases we can find a ``best case'' scenario, in which the effects of LV are maximal, as well as a ``worst case'' scenario, in which LV is present but is ineffective in the context of UHE neutrinos we are discussing now.

A fairly accurate general estimate of the minimum energy in which LV corrections in equation (\ref{eq:finaldisp}) is relevant is obtained, as stated previously, by comparing the largest mass of the particles entering in the LV reaction with the magnitude of the LV correction in these equations~\cite{Jacobson:2002hd}. In our case, assuming $\eta_{\nu I} \sim 1$, the typical energy at which LV contributions start to be relevant is of order $E_{th} \sim \sqrt{m_{\nu}\Mpl} \simeq 10^{13}~\eV$, assuming $m_\nu \simeq 10^{-2}~\eV$. This energy of 10 TeV allows observatories such as ICECUBE to possibly constrain these operators in the future using neutrino oscillations. We instead focus on changes to the UHE neutrino spectrum induced by these operators.

\section{UHE neutrinos and LV}
\label{sec:nuLV}

If we neglect exotic sources of UHE neutrinos (as suggested in many top-down models for the production of UHECRs, now disfavored by the current experimental photon limits), the ``cosmogenic'' neutrino flux is created \cite{Beresinsky:1969qj,Stecker:1973sy,Engel:2001hd,Semikoz:2003wv} via the decay of charged
pions produced by the interaction of primary nucleons with CMB
photons above $E_{pr} \simeq 5\times 10^{19}~\eV$, the
Greisen-Zatsepin-Kuz'min (GZK) effect \cite{gzk}.  HiRes \cite{Abbasi:2007sv} and AUGER \cite{Roth:2007in} spectral observations seem to confirm the presence of a GZK suppression in the UHECR spectrum. Although the suppression of the UHECR spectrum could be also due to the maximal accelerating power of UHECR sources, the fact that it occurs at just the right energy for being GZK taking place during propagation, and the results \cite{Cronin:2007zz} on the correlation of the UHECR arrival directions with the large scale distribution of matter within $\sim75~\Mpc$, seem to favor the GZK explanation. Motivated by these considerations, we make here the hypothesis that the GZK reaction is at work during UHECR propagation.

We have checked numerically that the process of pion decay is not strongly affected
by the above LV if only neutrinos are LV.  One might ask in general
if LV effects in the hadronic sector, which we have so far
neglected, can matter in the production of UHE neutrinos. We can
neglect the hadronic sector because of the tight constraints already
placed on such LV operators \cite{Galaverni:2007tq,Maccione:2008iw,Galaverni:2008yj,Maccione:2009ju}, which are stronger than what will be considered here.

Violation of Lorentz invariance however introduces new phenomena in the propagation of UHE neutrinos. A partial list of these effects includes:
\begin{description}
\item[Modified $\nu$ oscillations] Since the LV is parameterized in the mass eigenstate, the LV terms act as contributions to an effective mass and contribute to neutrino oscillations.  Cosmogenic neutrinos are not the right phenomena with which to study modified oscillations as they have oscillated many times during
their flight from an unknown source, making it extremely hard to derive
oscillation constraints.  ICECUBE and other neutrino observatories
may be sensitive to these operators for atmospheric neutrinos.
\item[$\nu$-\v{C}erenkov] \v{C}erenkov radiation in vacuum via charge-radius coupling or gravity: $\nu\rightarrow\nu\gamma$. This possibility has been already investigated for renormalizable operators~\cite{Coleman:1997xq,Coleman:1998ti}. The rate is too small for both renormalizable operators (at current limits) or non-renormalizable operators, even for cosmogenic neutrinos. It also involves additional LV coefficients for photon emission and perhaps new modes for graviton emission~\cite{Jacobson:2004ts}, thereby complicating the analysis and, at best, yielding less definitive limits.
\item[$\nu$-splitting]: $\nu\rightarrow  \nu \nu\nubar$.
This effect is what we focus on in the following, as it exclusively involves the neutrino sector and has a high enough rate to be seen at UH energies.
\item[$\nu$-pair emission]: $\nu\rightarrow \nu Z\rightarrow \nu
f\bar{f}$.  Here, $f$ represents some other fermion species besides
neutrinos.  Electrons are the only fermion species light enough to
have an effect compared to $\nu$-splitting at UH energies. We
initially will ignore this reaction, but we will consider the
possible effects of electron pair and hadronic emission in
section~\ref{sec:otherprocesses}.

\end{description}

\section{Neutrino Splitting}
\label{sec:nusplitting}

To understand how a neutrino can split into three, it suffices to
calculate the threshold energy for this reaction to occur, which in a
Lorentz invariant scenario would be infinite.  Let us compute the
threshold for the decay $\nu_A(p) \rightarrow\nu_A(p') \nu_B(q)
\nubar_B(q')$, where $\nu_A$, $\nu_B$ are neutrinos of mass
$m_{A}$,$m_B$ and $p,p',q,q'$ are the momenta. In the threshold
configuration the momenta of the outgoing particles are all aligned
and parallel to the direction of the incoming neutrino momentum
\cite{Mattingly:2002ba}, hence we choose all momenta to be in the
$z$-direction. Let $x$,$y$ and $t$ be the fraction of initial momentum
carried respectively by the outgoing $\nu_A$, by $\nu_B$, and by
$\nubar_B$, $t = 1-x-y$ and $0<x,y,t <1$.  For a general dispersion relation with LV term $\eta^{(n)}p^n/\Mpl^{n-2}$ the threshold equation (the energy conservation equation imposing momentum conservation) can then be written as
\begin{equation}
\frac{p^{n}}{\Mpl^{n-2}}\left[\eta_{\nu_A}^{(n)}(1-x^{n-1})-\eta_{\nu_B}(y^{n-1}+t^{n-1})\right]
= m_{\nu_A}^{2}\frac{1-x}{x} + m_{\nu_B}^{2}\left(\frac{1}{y} +
\frac{1}{t}\right)~. 
\label{eq:genthres}
\end{equation}

When LI is exact, the left hand side vanishes and there is no solution. With
LV, in the simple case where $n=4$ and $A=B$, we obtain the equation
\begin{equation}
\frac{p^{4}}{\Mpl^{2}m_{\nu_A}^{2}}\eta_{\nu_A}^{(4)} =
\frac{1}{3xyt}~. 
\label{eq:nuthreseq4}
\end{equation}

The minimum of the right-hand side in the equation above is 9 and is
attained at $x = y = t = 1/3$, hence the threshold energy is given
by
\begin{equation}
p_{\rm th (4)}^{\nu\rightarrow\nu\nu\nubar} = \sqrt{\frac{3\Mpl
m_{\nu_A}}{\left(\eta_{\nu_A}^{(4)}\right)^{1/2}}} \simeq 20~\TeV
\sqrt{\frac{m_{\nu_A}}{10^{-2}~\eV\left(\eta_{\nu_A}^{(4)}\right)^{1/2}}}~.
\label{eq:nuthres4}
\end{equation}

Note that $\eta_{\nu_A}$ must be positive in order for there to be
splitting, i.e.~we must deal with superluminal neutrinos. If $A \neq B$  then the
threshold energy must be solved for numerically as the LV parameter and mass are generically different. However, the existence of a finite threshold can still be shown for appropriate values of $\eta_{\nu_A}$ and $\eta_{\nu_B}$.  In the latter case, the coefficients do not need to be all positive for there to be a threshold, however it is necessary that $\eta_{\nu_B}< \eta_{\nu_A} - K(p)$, where $K(p)$ is a positive number that depends on the incoming momentum. Thresholds for negative coefficients have been investigated previously~\cite{Jacobson:2002hd,Konopka:2002tt} and there are strong limits on the LV coefficients.\footnote{ We notice here that according to our eq.~(\ref{eq:genthres}) neutrino splitting is forbidden for superluminal neutrinos in the case $n=1$ studied e.g.~in \cite{Carmona:2000ze}. Although the model presented in \cite{Carmona:2000ze} does not have an EFT description, a simple threshold computation under the assumption of energy-momentum conservation and equal values of masses and LV parameters for neutrinos and antineutrinos shows that if neutrino splitting is possible also in that framework, values $\eta^{(1)}_{\nu}\sim -10^{-51}$ can be probed by $\sim10^{19}~\eV$ neutrinos.}  However, for UHE neutrinos it is more difficult to extract constraints, as we shall discuss in Section~\ref{sec:notsogood}.

\subsection{The best case scenario: flavor blind LV}

Neutrinos created in a flavor eigenstate are, of course, a
combination of mass eigenstates. In order to maximize the neutrino
splitting signal one would want \textit{all} mass eigenstates to
split (``flavor blind'' scenario). The best case scenario is therefore when $\eta_{\nu I}>0$ for
each $I$ and each $\eta_{\nu I}$ is of the same order of magnitude
(so all states begin to decay at the same energy). As our best case
scenario then we take $\eta_{\nu I}$ equal and positive for all $I$.

\subsubsection{Decay time computation}
\label{sec:computation}

Let us now consider, within the above ``flavor blind scenario", energies well above threshold so that neutrinos do effectively split, and the best case combination of the $\eta_{\nu I}$.
The splitting rate can be calculated from the neutrino width
\begin{equation}
 \Gamma_{tot,A} = \sum_{\rm B} \Gamma_{AB}\:,
\end{equation}
where the sum runs over the open splitting channels and
$\Gamma_{AB}$ represents the partial width for the channel $\nu_A
\rightarrow\nu_A \nu_B \nubar_B$.  $\Gamma_{AB}$ is simply
\begin{equation} 
\label{eq:partialwidth}
 \Gamma_{AB} = \frac{1}{2E_p}\times \int |\mathcal{M}_{AB}|^2 \times
 d\Phi_{AB}\:,
\end{equation}
where $\Phi_{ AB }$ represents the phase space of the final states
and $\mathcal{M}_{AB}$ the matrix element for the process.  The
dominant channel for neutrino splitting is via the tree level
neutral current interaction. The matrix element for this
interaction is
\begin{equation}
\mathcal{M}_{A B} \propto
\frac{g^{2}}{4\cos^{2}\theta_{w}}\bar{u}_A(p')\gamma^{\mu} P_{L}u_A
(p)\frac{g_{\mu\nu}}{r^{2}-M_{Z}^{2} + i
M_Z\Gamma_Z}\bar{u}_B(q')\gamma^{\nu}P_{L}v_B(q)~, 
\label{eq:Mdec}
\end{equation}
where $g$ is the charged current electro-weak coupling constant, $P_{L}$ is the usual spin projector, $\theta_{w}$ is the Weinberg angle and $r^{\mu}$ represent the 4-momentum components of the $Z$ boson. $u$ and $v$ are the
spinors associated with $\nu_A$ and $\nu_B$, their functional form can be found in~\cite{Mattingly:2008pw}.  $r^2$ is at most the order of $p^4/M_{Pl}^2$ and hence for any incoming momenta $p<10^{19.5}~\eV$ $r^2 \ll M_Z^2$.  After some brute force algebra, we end with
\begin{equation}
\mathcal{M}_{AB} \propto
\frac{g^{2}}{4M_{Z}^{2}\cos^{2}\theta_{w}}\sqrt{16E_{p}E_{p'}E_{q}E_{q'}}\times
F(\chi)~,
\label{eq:M}
\end{equation}
where $\chi$ is the angle between incoming and outgoing $\nu_A$
and $F(\chi)$ is a complicated function of $\chi$.

The phase space, $d\Phi_{AB}$ is given by
\begin{equation}
d\Phi_{AB} = (2\pi)^{4}\delta^{(4)}(p_{in}-\sum
p_{out})\frac{d^{3}p'}{(2\pi)^{3}2E_{p'}}
\frac{d^{3}q'}{(2\pi)^{3}2E_{q'}} \frac{d^{3}q}{(2\pi)^{3}2E_{q}}~.
\end{equation}

We integrate over $q'$ and we are left with
\begin{equation} 
\label{eq:phasespace}
d\Phi_{AB} =
\frac{1}{8(2\pi)^{5}}\delta(E_{p}-E_{p'}-E_{q}-E_{q'})\frac{d^{3}p'
d^{3}q}{E_{p'}E_{q}E_{q'}}\;.
\end{equation}

Substituting $M_{AB}$ and $d\Phi_{AB}$ back into
(\ref{eq:partialwidth}) we have
\begin{equation} 
\label{eq:partialwidth2}
 \Gamma_{AB} =  \frac {g^4} {16 (2 \pi)^5 M_Z^4 cos^4 \theta_w
 } \int \delta(E_p-E_p'-E_q-E_q') d^3p' d^3q F^2~.
\end{equation}

We now turn to estimating the size of the remaining integral well
above threshold, as UHE neutrinos are far in excess of the neutrino
splitting threshold around 20 TeV. Temporarily, we will set
$\eta_\nu=1$ and re-insert it at the end. We first assume that the
opening angle for all three neutrinos is roughly equal, as is true
over most of phase space. In this case, $F(\chi)$ can be
approximated as $(1-\cos\chi)$.  The opening angle $\chi$
vanishes as $M_{Pl}\rightarrow \infty$, and is small even for UHE
neutrinos. $F(\chi)$ then reduces to $F(\chi)= \chi^2/2$. Indeed, 
$\chi$ is given by the typical transverse momenta $p_\perp$
divided by the longitudinal momenta $p_\parallel$. The
characteristic size of $p_\perp$ can be estimated by energy
conservation, recognizing that when energy is conserved any
``excess'' energy from LV goes into the energy needed to create
transverse momenta.  Formally this implies that
\begin{equation}
\frac{p_\parallel^3} {\Mpl^2} \sim \frac {p_\perp^2} p_\parallel,
\end{equation}
i.e. $p_\perp \sim p_\parallel^2 / \Mpl$.   $p_\parallel$ itself for
any particle can range from almost 0 to almost the initial energy
$E_p$ well above threshold.  The available phase space volume for
the remaining outgoing particles can therefore be approximated
(although somewhat overestimated) as a cylinder with length $E_p$
and radius $E_p^2/ \Mpl$, which has a total volume of $\pi E_p^5/
\Mpl^2$. Similarly, $\chi$ can be estimated as $\chi \sim
p_\parallel/\Mpl \sim E_p/\Mpl$.  The $\delta$-function in energy
simply removes one factor of energy from our equation.  Putting
these expressions back into (\ref{eq:partialwidth2}) we find
\begin{equation}
 \Gamma_{AB} \sim  \frac {g^4} {16 (2 \pi)^5 M_Z^4 cos^4 \theta_w}
 \frac {1} {E_p}
  \left(\frac {\pi E_p^5} {\Mpl^2}\right)^2 \frac {E_p^4} {4\Mpl^4}~,
\end{equation}
or, simplified
\begin{equation}
 \Gamma_{AB} \sim  \frac {G_F^{2}} {64 \pi^3
 \Mpl^8} E^{13}~,
\end{equation}
where we used $G_{F}/\sqrt{2} = g^{2}/(8M_{Z}^{2}\cos^{2}\theta_{w})$. 
If we now want to restore the $\eta_\nu$ factor, we notice that we
must put one $\eta_\nu$ for each $\Mpl^{-2}$. In the end, the total rate is computed by adding the partial rates for each neutrino flavor, which is equivalent to multiply by 3. Hence, we have
\begin{equation}
\Gamma_{\nu\nu\nubar} \sim
\frac{3G_{F}^{2}}{64\pi^{3}}\frac{\eta_{\nu}^{4}E^{13}}{\Mpl^{8}}\;.
\label{eq:gdecth}
\end{equation}
This width can be turned into a decay length as
\begin{equation}
L_{\nu\nu\nubar} = \frac{c}{\Gamma_{\nu\nu\nubar}} \sim 1.7\times
10^{-3}~\Mpc
~\eta_{\nu}^{-4}\left(\frac{E}{10^{19}~\eV}\right)^{-13}\;.
\label{eq:nudeclength}
\end{equation}

This makes clear we need to push the required energies above
$10^{18.5}~\eV$ (with $\eta_\nu = 1$) for the rate to be
appreciable. As a final remark, we notice that the decay length in Eq.~(\ref{eq:nudeclength}) strongly depends on both the energy and $\eta_{\nu}$. Therefore, the error about its actual magnitude we might have made in our estimate will reflect in very small errors in the determination of the energy at which LV effects start to be relevant as well as of the constraint on $\eta_{\nu}$.

\subsubsection{Z boson resonance}

At such high energies 
the Z could be real - i.e.~there is a resonance in the
matrix element. Even in this regime, however, the neutrino decay
time can be computed easily, as the only hypothesis one has to relax
is that the Z 4-momentum $r$ satisfies $r^2 \ll M_Z^2$.  The magnitude
of $r^2$ can be easily computed exploiting the kinematic equations.
We obtain $r^2 = 16/27\eta_\nu E_\nu^4/\Mpl^2$. The final decay
length is then
\begin{equation}
\fl
 L_{\nu\nu\nubar} \sim 1.7\times 10^{-3}~\Mpc ~\eta_{\nu}^{-4}\left(\frac{E_{\nu}}{10^{19}~\eV}\right)^{-13}\times\left[\left(1-\frac{16}{27}\eta_\nu\frac{E^4}{\Mpl^2 \,M_Z^2}\right)^2 + \left(\frac{\Gamma_Z}{M_{Z}}\right)^2\right]~.
 \label{eq:clength}
\end{equation}

\begin{figure}[htbp]
 \centering
 \includegraphics[scale=0.5]{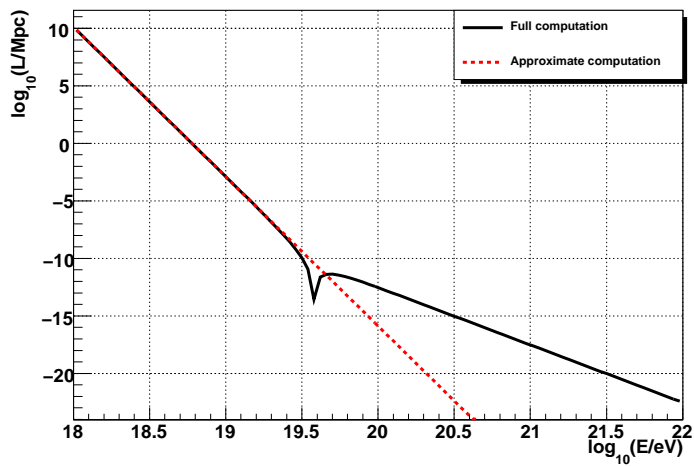}
 \caption{ Comparison between computations of the decay length without (Eq.~(\ref{eq:nudeclength})/red dashed line), and with (Eq.~(\ref{eq:clength})/black solid line) the Z boson resonance.}
 \label{fig:declength}
\end{figure}

A comparison between the two different decay lengths, Eq.~(\ref{eq:nudeclength}) and Eq.~(\ref{eq:clength}), can be found in
Fig.~\ref{fig:declength}. The Z resonance is hit at $E^4 =
27/16\,(\eta_\nu^{-1}\Mpl^2M_Z^2)$.
We notice that even though the two computations lead to very different results above the resonance, they will not lead to any appreciable effect in the neutrino spectra, as at such energies the decay lengths are anyway much smaller than the propagation distance of cosmological neutrinos.

\subsection{A non-optimal case: flavor dependent LV}

In the flavor blind case above every mass eigenstate undergoes splitting. It
is easy to construct a scenario where only one mass eigenstate
decays - one eigenstate has a positive $\eta$ LV coefficient while
the other two have no LV.  While this is unnatural in some sense, it
serves as a nice example to illuminate a method to easily hide LV.

The neutrino spectrum produced by the GZK process has a distinct
distribution in flavor states of 1/3 $\nu_e$ and 2/3 $\nu_\mu$. 
Therefore, in order to calculate a spectrum with LV
we need to convert this flavor spectrum into a mass spectrum, i.e.
we need to choose a neutrino mixing model. For our purposes we
choose tribimaximal mixing~\cite{Harrison:2002er,pdg}, which satisfies current
experimental constraints and provides a simple mixing matrix.  With
tribimaximal mixing the GZK neutrino spectrum has equal distribution
over all mass eigenstates.  

Now that we have spelled out out test models we shall proceed discussing the results for both scenarios.

\section{Results}
\label{sec:bestcase}

The net effect of neutrino splitting is very simple. It kills one
neutrino of energy $E$ and creates 3 neutrinos of average energy
$E/3$, provided the parent neutrino is above threshold and has a
reasonable life time. As we showed in sec.~\ref{sec:computation}, for a life time shorter than the age of the Universe, the neutrino energy has to be above $10^{18.5}~\eV$, i.e.~we need to probe UHE neutrinos.

The effects of neutrino splitting on the UHE neutrino spectrum are
twofold and can be understood qualitatively as follows.
\begin{description}
\item[Flux suppression at UH energies] The splitting is effectively an energy loss process for UHE neutrinos. If the rate is sufficiently high, the energy loss length can be below 1 Mpc. Let us call $\bar{E}(\eta_\nu)$ the energy at which this happens. Then, being GZK neutrinos produced mainly at distances larger than 1 Mpc, we do not expect any neutrino to be detected at Earth with $E>\bar{E}$.

The mere observation of neutrinos up to a certain energy $E_{\rm obs}$ would
imply a constraint, according to Eq.~\ref{eq:nudeclength}
\begin{equation}
\eta_{\nu}^{(4)} \lesssim
\left(\frac{E_{\rm obs}}{6\times10^{18}~\eV}\right)^{-13/4}\;.
\label{eq:constraint_naive}
\end{equation}

\item[Flux enhancement at sub-UH energies] Neutrinos lose energy by producing lower energy neutrinos. Eventually these neutrinos will become stable, either because their energy is below threshold, or because their lifetime is larger than their propagation time. Accordingly, we expect an enhancement of the neutrino flux at energies below ${\rm few} \times 10^{18}~\eV$. 
\end{description}

In spite of being qualitatively straightforward, however, this analysis is not powerful enough to provide us with constraints on LV. In order to obtain meaningful constraints, we have to resort to full MonteCarlo simulations of the UHECR propagation from sources to the Earth.

We simulated then the propagation of UHECR protons in the Inter Galactic Medium using the Monte Carlo package CRPropa~\cite{Armengaud:2006fx}, suitably modified to take into account LV in the neutrino sector.  The simulation parameters
are the following: we simulated unidimensional UHECR proton
propagation, with source energy spectrum $dN/dE \propto E^{-2.2}$,
from a spatially uniform distribution of sources located at redshift
$z < 3$ according to the Waxman \& Bahcall (WB) distribution used in \cite{Bahcall:2002wi}. The injection proton spectrum was tuned to fit AUGER data
\cite{Roth:2007in}.

\subsection{Results for flavor bind LV scenario}

\begin{figure}[htbp]
\centering
\includegraphics[scale = 0.7]{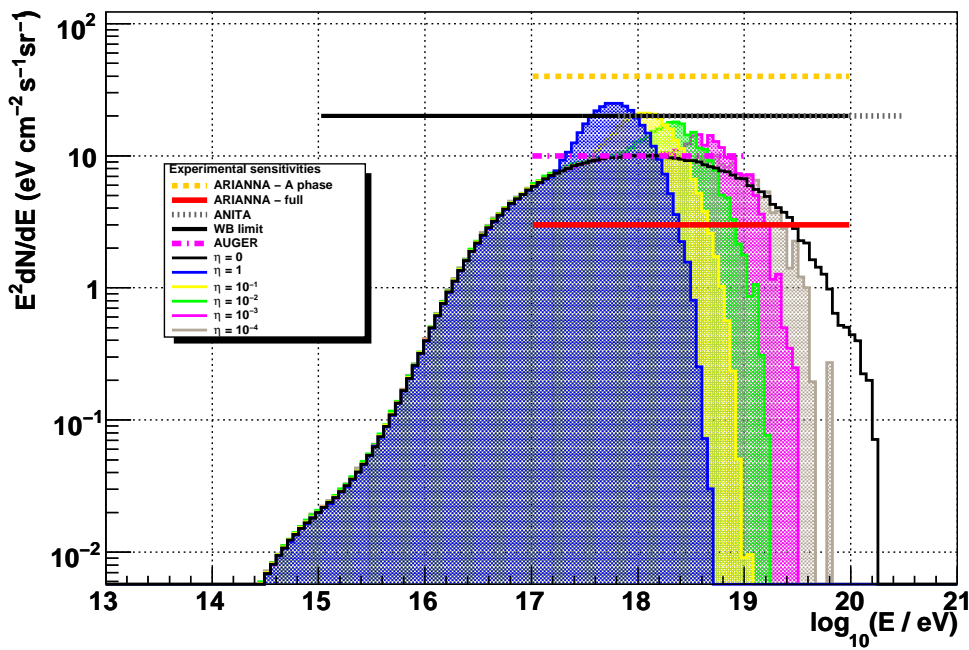}
\caption{Evolution of the predicted LV neutrino spectra varying
$\eta_\nu$ in the ``best case scenario". Sensitivities of main UHE neutrino operating and planned
experiments are shown, as found in
\cite{Barwick:2009zz,Gorham:2008yk,Collaboration:2009uy}. The Waxman \& Bahcall
limit \cite{Waxman:1998yy,Bahcall:1999yr} in the interesting energy
range is shown for reference.} \label{fig:spectra_eta}
\end{figure}

Figure~\ref{fig:spectra_eta} shows the outcome of the simulations for different values of the LV parameter $\eta_\nu$ in the best case scenario, together with experimental sensitivities from some existing and planned observatories, as well as the Waxman \& Bahcall bound \cite{Waxman:1998yy,Bahcall:1999yr} for reference.\footnote{ This limit is in fact an estimate of the neutrino luminosity of sources of UHE Cosmic Rays and $\gamma$-rays, in the hypothesis that the sources are optically thin to the escape of UHE particles and that both $\gamma$-rays and neutrinos are originated from UHECR interactions with radiation backgrounds. It is worth mentioning that this bound might be strongly affected by QG effects, as shown in \cite{AmelinoCamelia:2003nj}.}
Results are in agreement with qualitative expectations previously discussed: Above a
certain energy the neutrino spectrum displays a sharp cut off, while
at lower energy a peak appears. The peak position and strength, as
well as the position of the cut off, depend on the value of the LV
parameter $\eta_\nu$. If $\eta_\nu = 1$ the peak is found at
$\lesssim10^{18}~\eV$ and overwhelms the LI spectrum by a factor of
roughly 10, while the cut off energy is at $\sim 10^{18.7}~\eV$.
Also the evolution of the neutrino spectra with $\eta_\nu$ can be
seen in the same Fig.~\ref{fig:spectra_eta}.

Although at present it is not possible to draw firm conclusions for the constraints on LV in the neutrino sector, we notice that future experiments, if not already AUGER, will be able to probe the fluxes we predict, hence they will be able to cast constraints on
$\eta_\nu$. In the future, the ARIANNA experiment \cite{Barwick:2006tg} will be able to probe fluxes down to $\sim3~\eV\,\cm^{-2}\,\s^{-1}\,\sr^{-1}$ in the energy range $10^{17} \div 10^{20}~\eV$ in the full configuration \cite{Barwick:2009zz}. If such a sensitivity will be achieved, a constraint of order $\eta_{\nu} \lesssim 10^{-4}$ will be cast, according to Fig.~\ref{fig:spectra_eta}.
Moreover, the IceRay experiment \cite{Allison:2009rz} planned at the South pole is expected to observe roughly 4 neutrino events per year of data taking for the WB model preferentially studied here, in the range $10^{17} \div 10^{19.5}~\eV$ \cite{Allison:2009rz}. Hence, constraints of order $\eta \lesssim 10^{-3}$ are expected after few years of data taking by this experiment.
We finally notice here that by exploiting this strategy nothing can be said
about the case $\eta_\nu < 0$.

\subsection{Results for flavor dependent LV scenario}
\label{sec:notsogood}

\begin{figure}[htbp]
\centering
\includegraphics[scale = 0.7]{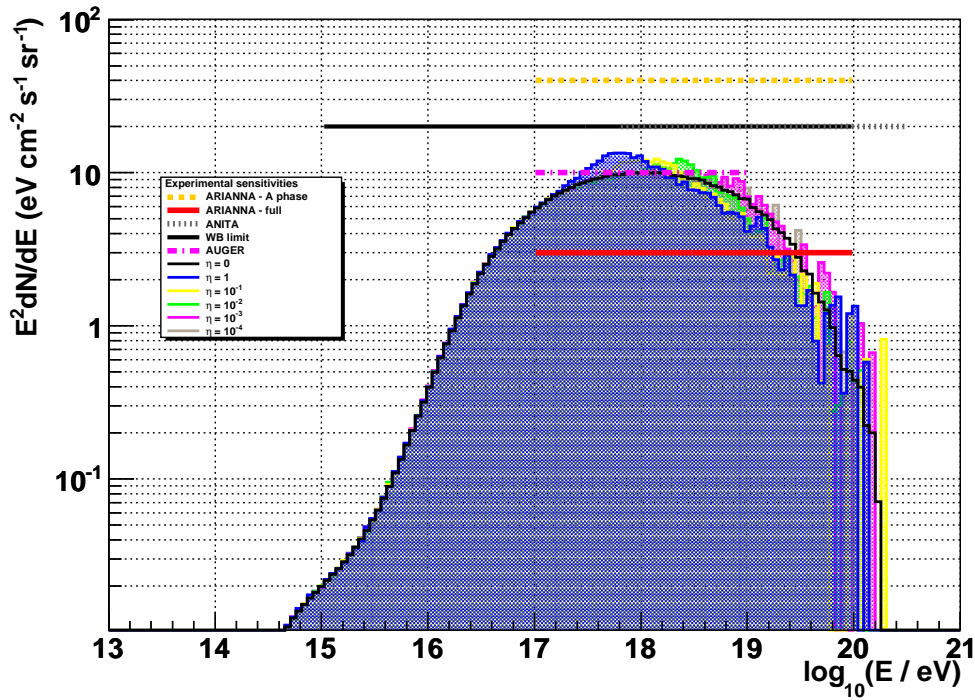}
\caption{Observed UHE neutrino spectra for different values of the LV coefficient of the only mass eigenstate undergoing splitting in the non-optimal scenario.
The flux suppression at UH energies has effectively disappeared (for what concerns observational relevance) and the excess at lower energies is significant only for $O(1)$ LV coefficient.} 
\label{fig:onemasssplits}
\end{figure}

If only one mass eigenstate decays, the energy momentum conservation equations, rates, and lifetimes all remain the same. 
We therefore can apply our MonteCarlo spectrum for this scenario to just any one of the mass eigenstates (given our assumption of tribimaximal mixing no neutrino mass eigenstate is preferred).
The only difference is that only 1/3 of the total flux undergoes neutrino splitting. 
The resulting spectrum is shown in Fig.~\ref{fig:onemasssplits}. Note that even for $\eta=1$ the departure from the LI spectrum is significantly reduced, as one would expect, with the magnitude of the deviation at the peak being lowered to a $\sim 50\%$ excess over the LI flux. Hence, with present sensitivity, absence of a signal in the neutrino spectrum does not mean that LV is ruled out for neutrinos, even for $O(1)$ positive coefficients.
We notice that the cutoff feature present in the flavor blind case has now effectively disappeared. Then, observations of a UHE neutrino flux up to some maximal energy do not allow to draw conclusions about LV in this case scenario. 

We now return to the question of negative coefficients and neutrino splitting. So far, in both our flavor blind and flavor dependent scenarios all the coefficients have been positive. As argued in section~\ref{sec:nusplitting}, neutrinos can split even with negative coefficients when the relation $\eta_{\nu_B}< \eta_{\nu_A} - K(p)$ holds. However, one can immediately see that therefore at least one of the mass eigenstates is always stable if the coefficients are negative. This implies that no equivalent of the ``best case scenario" (all the mass eigenstates decay) is possible for negative $\eta_{\nu I}$ and that, consequently, the deviation from the LI spectrum will be suppressed by at least 1/3. Therefore cosmogenic neutrinos are not effective probes of the $\eta_{\nu I}<0$ region of parameter space.

\subsection{Dependence on model uncertainties}

Predictions of cosmogenic neutrino spectra are known to be plagued by several uncertainties, as they depend strongly on the evolution of UHECR sources with redshift, which cannot be directly probed with UHECRs due to the GZK attenuation effect. As a consequence, also our predictions for the UHE neutrino flux in the presence of LV can be affected.

\begin{figure}[htbp]
\centering
\includegraphics[scale = 0.7]{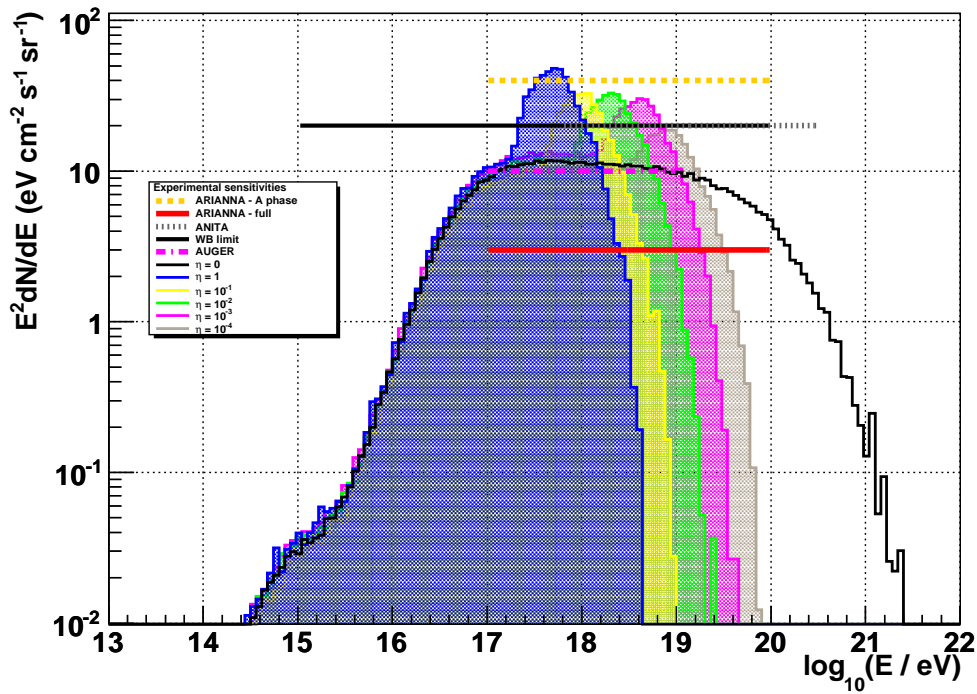}
\caption{Evolution of the predicted LV neutrino spectra varying
$\eta_\nu$ for the maximal model \cite{Semikoz:2003wv}.}
\label{fig:max}
\end{figure}

We check this by simulating different UHECR sources evolution models for the ``best case scenario". In Fig.~\ref{fig:max} we show the results for the model outlined in \cite{Semikoz:2003wv}, which is expected to give the largest flux.
As a result, the presence and the strength of the bump feature are demonstrated to be weakly dependent on the underlying UHECR source distribution, with the choice of the source distribution affecting by a factor of 2 the height of the bump. This was somewhat expected from our qualitative considerations: as long as UHECR sources are located at distances much larger than 1 Mpc from Earth, the magnitude of the bump depends only on how many neutrinos were originally produced at energy larger than the LV cutoff energy. On the other hand, the cutoff feature is definitely model independent, as long as the unknown maximal energy at which UHECRs are accelerated is enough to produce neutrinos at energy larger than $\sim 6\times10^{18}\cdot \eta^{-4/13}~\eV$.

\section{Other processes} \label{sec:otherprocesses}

As we mentioned in the previous sections, there are other processes,
besides splitting, leading to neutrinos losing energy in the
inter-galactic medium. In particular, the hadronic decay modes of
the Z$^{0}$ could lead to the production of UHE protons, thereby
mimicking a ``Z-burst'' effect \cite{Fargion:1997ft}. Although this
model was conceived for other purposes than constraining LV, and has
now lost most of its attraction due to experimental results on proton
spectra above the GZK threshold, it is notable that the same
mechanism could in principle help constrain LV in the neutrino
sector. However, we can argue with a very simple argument that this
is not the case.

\begin{figure}[htbp]
\centering
\includegraphics[scale = 0.7]{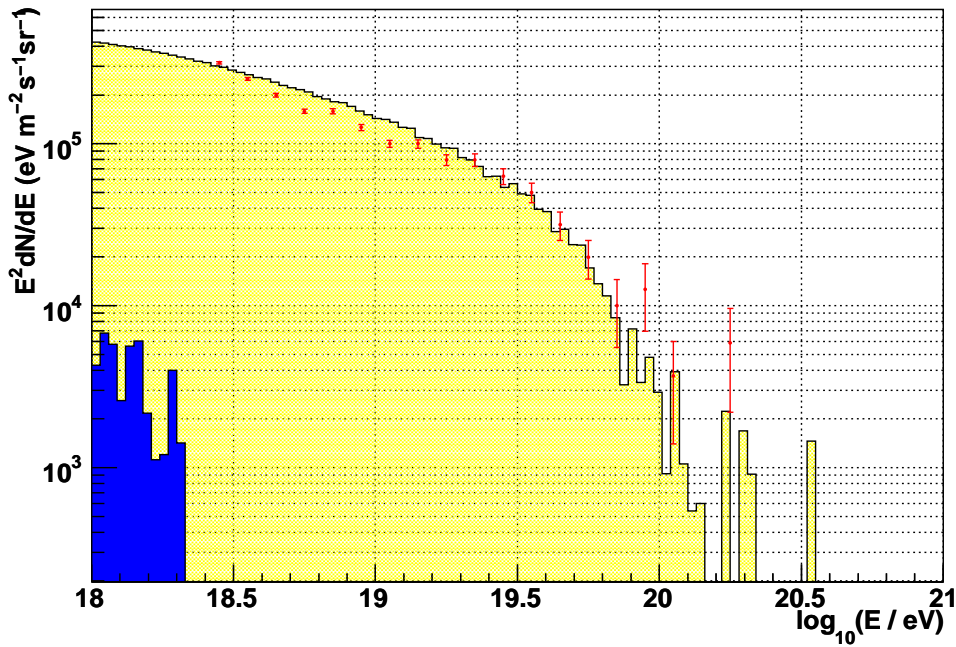}
\caption{Observed UHE proton spectrum compared with predictions for
just primary (yellow) and ``Z-burst''-like originated protons (blue). The latter might produce an excess in the UHECR spectrum at $10^{18}~\eV$, no larger than 5\%, and hence below the systematic uncertainties of UHECR experiments.
Note that, in order to speed up computation, we switched off pair production, which is known to
play an important r\^ole in reproducing the ankle feature \cite{Aloisio:2006wv}. This is why our UHECR spectrum is above data between $10^{18.5}~\eV$
and $10^{19}~\eV$.} \label{fig:zburst}
\end{figure}

Let us consider the LI neutrino flux, which corresponds in
Fig.~\ref{fig:spectra_eta} to the $\eta=0$ case. 
The minimal process leading to proton
production in this context is $\nu\rightarrow \nu Z^{*}\rightarrow
\nu p \bar{p}$, whose threshold energy, if $\eta_\nu = 1$, is at
$\sim10^{18.7}~\eV$. Processes involving more particles are
naturally suppressed, however they represent the majority of the allowed channels and hence they might be relevant to our
case. Even if we do not know the energy spectrum of the produced protons,
we know that their mean energy is roughly 1/80 times the energy of
the parent neutrino \cite{Fargion:1997ft}. We can then compute the
maximal expected flux of UHECRs produced by this Z-burst-like
mechanism by converting all the neutrinos with energy $E_{\nu}$
above threshold into 2 protons of energy $1/80\times E_{\nu}$. The
result is shown in Fig.~\ref{fig:zburst}. 
This mechanism would
indeed contribute to the UHECR spectrum at $10^{18}~\eV$, producing
an excess no larger than 5\%, which however is below the systematic
uncertainties of UHECR experiments. Therefore, no constraints on LV
can be implied using this technique.

Electron/positron pair production would also contribute to the
cutoff in the neutrino spectrum.  The threshold for pair production
of electrons is at roughly $\sqrt{m_e \Mpl}=10^{17}$ eV.  At
energies near the existing neutrino splitting cutoff, $10^{18.5}$ eV
we are well above threshold.  The mass of the electrons is therefore
irrelevant in calculating the decay rate (just as the neutrino mass
was) and hence the electron pair production process has nearly the
same rate as the neutrino splitting process.  If neutrinos produce
electron/positron pairs (which after all depends on LV in the
electron sector) it can only increase the steepness of the cutoff.
It will not contribute to the excess neutrino flux at $10^{17.5}$ eV
and hence one needs to probe energies one order of magnitude larger than $10^{17.5}~\eV$ to derive constraints exploiting the presence or the absence of a cutoff feature.

\section{Conclusions}
\label{sec:conclusions}

In this work we have investigated possible signals of higher dimension LV CPT even operators in the UHE neutrino spectrum, in particular the effect of the neutrino splitting on the UHE neutrino spectrum. This process provides a clean test as it does not involve other LV operators apart from the neutrinos' ones.  In addition, since the dominant neutrino field is left-handed, there are no complications due to different LV for different chirality of fermion, as there are in the UHECR case, and one ends up with one LV coefficient for each neutrino mass eigenstate.
In the flavor blind scenario, where every mass eigenstate undergoes splitting approximately at the same energy, there is both a precocious fall off of the neutrino flux at UHE as well as a significant excess in the UHE neutrino flux at energies as low as $10^{17.5}$ eV. Noticeably, this kind of energies are well within reach of current and future UHECR experiments and about order of magnitude below those so far used for LV tests with hadronic and electromagnetic UHECRs.

According to our study, existing or planned UHE neutrino experiments have the potential to probe LV in the neutrino sector for coefficients $\eta \gtrsim 10^{-4}$. However, we have discovered a serious difficulty with deriving constraints in the absence of a positive LV signal event. The distribution in mass eigenstates is roughly equal for UHE neutrinos in realistic mixing scenarios. Although it might seem somewhat unnatural that LV has different effects on different mass states of the same particle field, if this is the case then it is possible that LV can exist/be strong for one mass eigenstate yet be almost invisible for UHE neutrino detectors. In particular, the observation of a neutrino flux up to some maximal energy does not imply a firm conclusion on LV, at least with present accuracy. However, let us note that if a neutrino is ever detected at energy $E_{\nu}$, we can rule out a flavor blind $\eta_{\nu} \gtrsim (E_{\nu}/10^{18.8}~\eV)^{-13/4}$ according to Eq.~(\ref{eq:constraint_naive}).
On the other hand, the bump feature can lead to constraints on $\eta_{\nu} > 0$, as this bump should be observable in any such scenario of LV in the neutrino sector. In this case, to obtain a $O(1)$ constraint on LV requires at least a 50\% accuracy in the determination of the neutrino spectrum in the energy range $10^{17}\div10^{19}~\eV$, which can be achieved by future experiments.

\ack
We thank J.~Kelley and B.~McElrath for useful discussions. This work was supported by the Deutsche Forschungsgemeinschaft through the collaborative research centre SFB 676 ``Particles, Strings and the Early Universe: The Structure of Matter and Space-Time''. LM acknowledges support from the State of Hamburg, through the Collaborative Research program ``Connecting Particles with the Cosmos'' within the framework of the LandesExzellenzInitiative (LEXI). 

\section*{References}

\end{document}